\documentclass{PoS}
\usepackage{natbib}

\newcommand{\fermilat}{\textit{Fermi}-LAT}
\newcommand{\srcs}{QSO B0218+357}
\newcommand{\srclens}{[PBK93]\,B0218+357\,G}

\title{Detection of very-high-energy gamma rays from the most distant and gravitationally lensed blazar \srcs\ using the MAGIC telescope system}

\ShortTitle{Detection of \srcs\ with the MAGIC telescopes}

\author{\speaker{Julian Sitarek}\\
        University of \L\'od\'z, PL-90236 Lodz, Poland, \\
        IFAE, Campus UAB, E-08193 Bellaterra, Spain
        E-mail: \email{jsitarek@uni.lodz.pl}}

\author{J.~Becerra Gonz\'alez\\
        Inst. de Astrof\'isica de Canarias, E-38200 La Laguna, Tenerife, Spain; Universidad de La Laguna, Dpto. Astrof\'isica, E-38206 La Laguna, Tenerife, Spain\\
now at NASA Goddard Space Flight Center, Greenbelt, MD 20771, USA and Department of Physics and Department of Astronomy, University of Maryland, College Park, MD 20742, USA
        E-mail: \email{jbecerragonzalez@gmail.com}}

\author{Dijana Dominis Prester\\
        Croatian MAGIC Consortium, Rudjer Boskovic Institute, University of Rijeka and University of Split, HR-10000 Zagreb, Croatia
        E-mail: \email{dijana@phy.uniri.hr}}

\author{Elina Lindfors\\
        Tuorla Observatory, Department of Physics and Astronomy, University of Turku, Finland
        E-mail: \email{elilin@utu.fi}}

\author{Marina Manganaro\\
Inst. de Astrof\'isica de Canarias, E-38200 La Laguna, Tenerife, Spain; Universidad de La Laguna, Dpto. Astrof\'isica, E-38206 La Laguna, Tenerife, Spain
        E-mail: \email{manganaro@iac.es}}

\author{Daniel Mazin\\
        ICRR, The University of Tokyo
        E-mail: \email{mazin@icrr.u-tokyo.ac.jp}}

\author{Miguel Nievas Rosillo\\
        Universidad Complutense, E-28040 Madrid, Spain
        E-mail: \email{miguelnievas@ucm.es}}

\author{Antonio Stamerra\\
        INAF National Institute for Astrophysics, I-00136 Rome, Italy
        E-mail: \email{stamerra@oato.inaf.it}}

\author{Ievgen Vovk\\
        Max-Planck-Institut f\"ur Physik, D-80805 M\"unchen, Germany
        E-mail: \email{Ievgen.Vovk@mpp.mpg.de}}

\author{for the MAGIC Collaboration}

\author{Sara Buson\\
  Center for Research and Exploration in Space Science and Technology (CRESST)\\
  NASA Goddard Space Flight Center, Greenbelt, MD 20771, USA\\
  University of Maryland Baltimore County, Baltimore, MD 21250, USA\\
  E-mail: \email{sara.buson@nasa.gov}}

\author{on behalf of the Fermi-LAT Collaboration}

\abstract{
\srcs\ is a blazar located at a cosmological redshift of z=0.944.
It is gravitationally lensed by a spiral galaxy at a redshift of z=0.68.
The blazar and its lens are well studied in the radio through X-ray bands, and several blazar outbursts were detected by \fermilat\ at energies above 100 MeV.
Strong gravitational lensing was invoked to explain the two components apparent in the radio and GeV light curves, separated by 10-12 days.
In July 2014 another outburst was observed by \fermilat , triggering follow-up observations with the MAGIC telescopes at energies above 100 GeV.
The observations were scheduled at the expected time of arrival of the component delayed by the strong gravitational field of the lens, resulting in a firm detection of \srcs .
Using the combined \fermilat\ and MAGIC data sets, we report on variability of this unique blazar, the most distant among all currently known very high energy sources.
}

\FullConference{The 34th International Cosmic Ray Conference,\\
		30 July- 6 August, 2015\\
		The Hague, The Netherlands}

\begin{document}
\section{Introduction}
Even while there are already over 50 blazars\footnote{http://tevcat.uchicago.edu} detected in the very high energy (VHE $\gtrsim100\,$GeV) range, most of them are relatively close-by sources with redshift $z\lesssim0.5$.
Until recently, the farthest sources observed in this energy range have been: 3C\,279 ($z=0.536$, \citealp{al08}), KUV\,00311-1938 ($z>0.506$, \citealp{be12}) and PKS1424+240 ($z>0.6$, \citealp{acc10}).
Observations of farther sources in VHE gamma-rays are difficult due to the strong absorption in the interaction with the extragalactic background light (EBL), \cite{gs66}.
At the redshift of $\sim 1$ it results in a cut-off at the energy\footnote{The energy throughout the text is given in the Earth's frame of reference} of $\sim 100\,$GeV.
Such energies are at the lower edge of the performance of the current generation of Imaging Atmospheric Cherenkov Telescopes (IACTs), making such observations challenging.
To maximize the detection chance, the observations are often triggered by a high state observed in lower energy ranges. 
In particular, \fermilat\ scanning the whole sky in GeV range can provide alerts of high energy fluxes and spectral shape. 
Unfortunately due to required integration and processing time and very limited duty cycle of IACTs the inevitable delay of the follow-up observation may exceed the duration of the flare.

\srcs , also known as S3\,0218+35, is a blazar, most probably a flat spectrum radio quasar (FSRQ), located at the redshift of $0.944$ \citep{li12}.
The object is gravitationally lensed by a galaxy \srclens\ located at the redshift of $0.68$\footnote{https://www.cfa.harvard.edu/castles/}.
The radio image shows two distinct components with the angular separation of only 335\,mas and an Einstein's ring \citep{od92}.
Observations of variability of the two radio components led to a measurement of a delay of 10-12 days \citep{co96,bi99}.
The delayed component had a 3.57-3.73 times weaker flux \citep{bi99}, however at lower radio frequencies the ratio decreases to $\sim 2.6$ \citep{mi06}.
The difference can be explained to occur due to additional free-free absorption in the line of sight to one of the radio images  \citep{mi07}. 
\srcs\ is one of only two objects with a measured gravitational lensing effect in GeV energy range. 
In 2012 it went through a series of outbursts registered by the \fermilat\ instrument \citep{ch14}. 
Even while \fermilat\ does not have the necessary angular resolution to disentangle the two emission components, the statistical analysis of light curve autocorrelation function led to a measurement of time delay of $11.46\pm0.16$ days. 
Interestingly the magnification factor, contrary to the radio measurements, was estimated to be $\sim1$.

Another flaring state of \srcs\ was observed by \fermilat\ in July 2014 \citep{atel6316}. 
The original flare could not be observed by MAGIC due to the full moon period.
Observations were scheduled in the expected time of the arrival according to the previously measured by \fermilat\ delay.
Those MAGIC observations resulted in the discovery of VHE gamma-ray emission from \srcs\ \citep{atel6349}.

MAGIC (Major Atmospheric Gamma Imaging Cherenkov) is a system of two Cherenkov telescopes located in the Canary Island of La Palma.
Due to the large 17m diameter mirror dishes the telescopes perform observations of gamma rays with energies $\gtrsim$50\,GeV.
Here we report the results of the observations by the MAGIC telescopes which led to the detection of \srcs\ during the flaring state in July 2014.

\section{MAGIC: observations and data analysis}
In summer 2012 MAGIC finished a major upgrade \citep{al15a} greatly enhancing the performance of the instrument \citep{al15b}.
The sensitivity of the MAGIC telescopes achieved in the best energy range ($\gtrsim300\,$GeV) is at the level of $\sim 0.6\%$ of Crab Nebula flux in 50\,h of observations (for a gamma-ray excess of 5 times the RMS of the residual cosmic-ray background).
The angular resolution of MAGIC is of the order of $0.08^\circ$, i.e. insufficient for disentangling the emission from the two components of \srcs .

The telescopes could not follow the original flare seen by \fermilat\ from \srcs\ in 2014 as it appeared during the full moon time, when observations with low energy threshold are not possible.
Following the previous measurements of the gravitational lensing delay observations were scheduled in the expected time of arrival of the second component. 
The observations  were performed in 14 nights between 23\textsuperscript{rd} of July and 5\textsuperscript{th} of August 2014 for a total duration of 12.8\,h at intermediate zenith angle ($20^\circ-43^\circ$).
The data reduction (stereo reconstruction, gamma/hadron separation and estimation of the energy and arrival direction of primary particle) was performed using the standard analysis chain of MAGIC \citep{magic_mars, al15b}.
As \srcs\ is an extragalactic source, it is affected by $\sim 30\%$ smaller night sky background than the Crab Nebula data used to estimate the telescope performance in \cite{al15b}. 
Therefore we were able to apply image cleaning thresholds lower by $\sim 15\%$ with respect to the ones used in the standard analysis presented in \citet{al15b}, lowering the energy threshold of the analysis. 
The analysis was performed using a dedicated Monte Carlo (MC) with night sky background and trigger parameters tuned to reproduce as accurately as possible the actual observation conditions.

\section{\fermilat : observations and data analysis}
LAT data collected between MJD 56849--56875 (2014 July 11\textsuperscript{th} -- August 8\textsuperscript{th}) were extracted from a circular region of interest (ROI) of $15^{\circ}$ radius centered at the \srcs\ radio position, ${\rm~R.A.}=35^\circ.27279$, ${\rm~Decl.}=35^\circ.93715$ \citep[J2000;][]{pat92} and analyzed in the energy range $0.3 - 300$ GeV using the standard \textit{Fermi Science Tools} (version {\tt v9r34p1}) in combination with the {\tt P7REP\_SOURCE\_V15} LAT Instrument Response Functions.
We applied the {\tt gtmktime} filter ($\#$3) to the LAT data following the FSSC recommendations\footnote{{\tt http://fermi.gsfc.nasa.gov/ssc/data/analysis/documentation/Cicerone/\\Cicerone\_Likelihood/Exposure.html}}. 
According to this prescription, time intervals when the LAT boresight was rocked with respect to the local zenith by more than $52^{\circ}$ (usually for calibration purposes or to point at specific sources) and events with zenith angle $>100^{\circ}$  were excluded to limit the contamination from Earth limb photons. 

The spectral model of the region included all sources located within the ROI with the spectral shapes and the initial parameters for the modeling set to those reported in the 3FGL \citep[][]{3fgl} as well as the isotropic  ({\tt iso\_source\_v05.txt}) and  Galactic diffuse  ({\tt gll\_iem\_v05\\.fit}) components.
In generating the light curve the source of interest was modeled with a power-law spectral shape with normalisation and index free to vary.
To access the detection significance we used the Test Statistic (TS) value.
The TS quantifies the probability of having a point gamma-ray source at the location specified and corresponds roughly to the square of the standard deviation assuming one degree of freedom \citep{mat96}.
It is defined as TS = $-2\log (L_0 /  L)$,  where $L_0$ is the maximum likelihood value for a model without an additional source (the 'null hypothesis') and $L$ is the maximum likelihood value for a model with the additional source at the specified location.
We note that in our analysis the second model had two more degree of freedom (i.e. normalisation and index were left free), therefore TS=9 (25) corresponds to significance of $\sim$2.5 (4.6) $\sigma$ respectively.
During the analysed period  \srcs\ was not always significantly detected. Flux upper limits at the 95\% confidence level were calculated for each interval where the source TS was $<9$.

\section{Results}
\begin{figure}[t]
\centering
\includegraphics[width=0.6\textwidth]{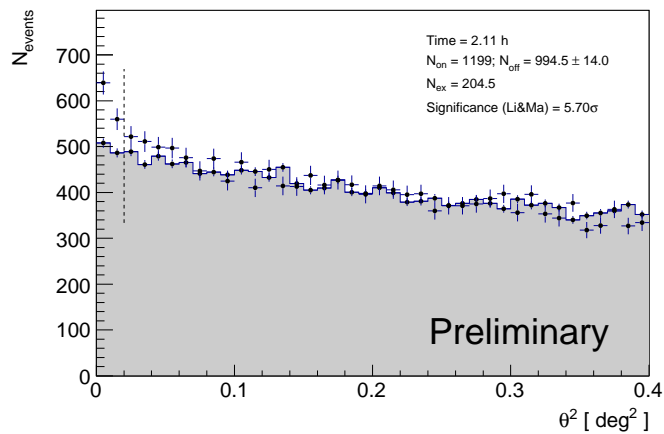}
\caption{Distribution of the squared angular distance, $\theta^2$, between the reconstructed source position and the nominal source position (points) or the background estimation position (shaded area). 
Vertical dashed line shows the value of $\theta^2$ up to which the number of excess events and significance are computed. 
}\label{fig:th2}
\end{figure}

The VHE gamma-ray emission was detected on the nights of 25\textsuperscript{th} and 26\textsuperscript{th} of July, during the expected delayed component of the \fermilat\ flare.
The cuts providing the best sensitivity in the 60-100 GeV range according to \citealp{al15b} were used for the source detection.
The total observation time during those 2 nights, 2.11\,hr, yielded a statistical significance of $5.7 \sigma$. (see Fig.~\ref{fig:th2}). 

\begin{figure}[t]
\centering
\includegraphics[width=0.9\textwidth]{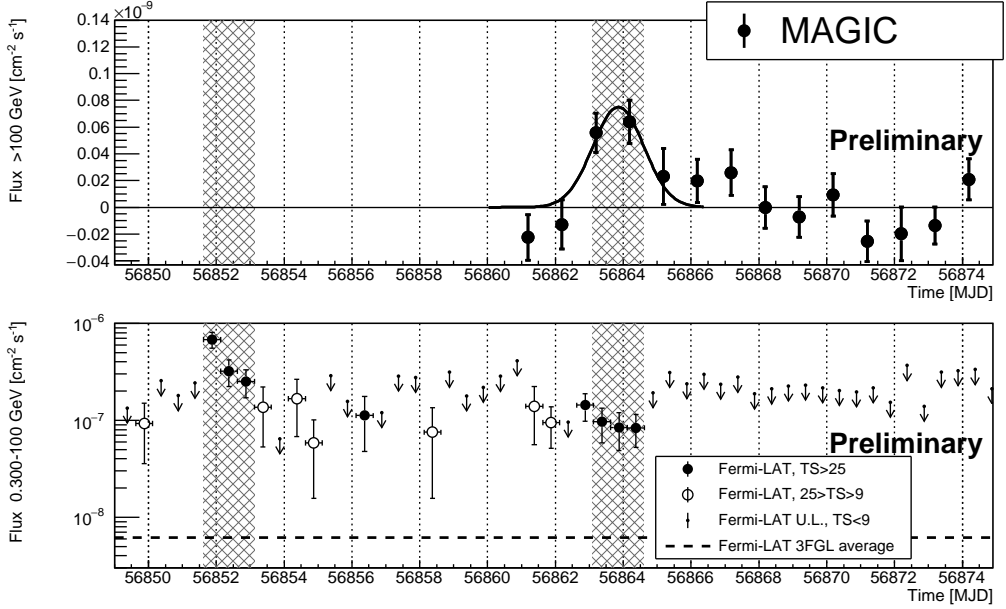}
\caption{
Light curve of \srcs\ during the flaring state in July/August 2014. 
Top panel: nightly MAGIC fluxes above 100\, GeV (points) and a Gaussian fit to the peak position (thick solid line). 
Bottom: \fermilat\ emission above 0.3\,GeV binned in 12\,h bins (plotted in log scale). 
The shaded regions (separated by 11.5 days) show the original and delayed flare following \citep{bu15}.
Vertical dashed line is the average flux of \srcs\ above 0.3\,GeV \citep{3fgl} 
}\label{fig:lc}
\end{figure}

The light curve obtained by MAGIC above 100\,GeV is compared with the emission observed above 0.3\,GeV by \fermilat\ in Fig.~\ref{fig:lc}.
The \fermilat\ emission during the expected delayed component of the flare is a factor 20 higher than the average state of the source \citep{3fgl}. 
During both components of the flare the photon index was significantly harder \citep{bu15} than the average one \citep{3fgl}. 
A fit with a simple Gaussian function gives the characteristic time scale of the MAGIC flare below 1 day. 
The two flaring nights give the mean flux of $(5.8 \pm 1.6_{\rm stat} \pm 2.4_{\rm syst})\times 10^{-11}\mathrm{cm^{-2}\,s^{-1}}$ above 100\,GeV.
The relatively large systematic error is a result mainly of a 15\% uncertainty in the energy scale. 

\section{Discussion and conclusions}
The MAGIC telescopes has detected VHE gamma-ray emission from gravitationally lensed blazar, \srcs , during the second component of a flare in July 2014. 
The previously known delay made it possible to schedule the observations in advance, allowing to see also the raise of the flare. 
\srcs\ and, detected a few months later, PKS1441+25 \citep{atel7416, atel7433}, are currently the two most distant sources (redshift $z\sim 0.9$) known to emit in VHE gamma-rays. 
Moreover \srcs\ is the only gravitationally lensed blazar detected at the VHE energies. 
The spectral features of the emission, modelling and impact on the EBL measurements will be discussed in future work. 

\section{Acknowledgments}

We would like to thank
the Instituto de Astrof\'{\i}sica de Canarias
for the excellent working conditions
at the Observatorio del Roque de los Muchachos in La Palma.
The financial support of the German BMBF and MPG,
the Italian INFN and INAF,
the Swiss National Fund SNF,
the ERDF under the Spanish MINECO (FPA2012-39502), and
the Japanese JSPS and MEXT
is gratefully acknowledged.
This work was also supported
by the Centro de Excelencia Severo Ochoa SEV-2012-0234, CPAN CSD2007-00042, and MultiDark CSD2009-00064 projects of the Spanish Consolider-Ingenio 2010 programme,
by grant 268740 of the Academy of Finland,
by the Croatian Science Foundation (HrZZ) Project 09/176 and the University of Rijeka Project 13.12.1.3.02,
by the DFG Collaborative Research Centers SFB823/C4 and SFB876/C3,
and by the Polish MNiSzW grant 745/N-HESS-MAGIC/2010/0.
JS is supported by Fundacja U\L .

The \textit{Fermi}-LAT Collaboration acknowledges support for LAT development, operation and data analysis from NASA and DOE (United States), CEA/Irfu and IN2P3/CNRS (France), ASI and INFN (Italy), MEXT, KEK, and JAXA (Japan), and the K.A.~Wallenberg Foundation, the Swedish Research Council and the National Space Board (Sweden). Science analysis support in the operations phase from INAF (Italy) and CNES (France) is also gratefully acknowledged.

\end{document}